\documentclass[aps,preprint,superscriptaddress]{revtex4-1}

\usepackage{amsmath}    
\usepackage{graphicx}   
\usepackage{color}      
\usepackage{hyperref}   
\usepackage{todonotes}
\usepackage{nicefrac}

\usepackage[T1]{fontenc}
\usepackage[utf8]{inputenc}

\begin{document}

\title{Electronic excitation spectrum of doped organic thin films investigated using electron energy-loss spectroscopy}
\author{Friedrich Roth}
\affiliation{Center for Free-Electron Laser Science / DESY, Notkestra\ss e 85, D-22607 Hamburg, Germany}

\author{Martin Knupfer}
\affiliation{IFW Dresden, P.O. Box 270116, D-01171 Dresden, Germany}

\date{\today}

\begin{abstract}
The electronic excitation spectra of undoped, and potassium as well as calcium doped phenantrene-type hydrocarbons have been investigated using electron energy-loss spectroscopy (EELS) in transmission. In the undoped materials, the lowest energy excitations are excitons with a relatively high binding energy. These excitons also are rather localized as revealed by their vanishing dispersion. Upon doping, new low energy excitation features appear in the former gaps of the materials under investigation. In K$_3$picene and K$_3$chrysene they are characterized by a negative dispersion while in Ca$_3$picene they are dispersionless.
\end{abstract}

\maketitle

\section{Introduction}

The physics of molecular solids is as rich as that of other materials. For instance, various electronic ground states can be achieved which cover the full range from insulators to metals and superconductors \cite{Toyota2007,Quahab2004,Pope1999,Gunnarsson1997}. Aromatic hydrocarbons are prominent building blocks in the class of molecular solids. Their $\pi$ electron system results in semiconducting properties, and these materials combine the advantages of their relatively low cost with the possibility of modifying them using the methods of synthetic organic chemistry in a practically unlimited fashion. Their potential application in organic electronic devices has motivated many investigations in the past. These have, for instance, been exploited in organic field effect transistors in view of fundamental as well as applied aspects \cite{Horowitz2004,Clemens2004,Scheinert2004}. Moreover, organic semiconductors are also of interest for the manufacture of organic photovoltaic cells, organic light emitting diodes or organic spintronics \cite{Forrest2004,Murphy2007,Mullen2006}.

\par

The ability to incorporate electron acceptors and donors into the relatively open crystal structure of molecular solids enables the control of their electronic properties by introducing charge carriers. This special aptitude represents a promising route for technology as well as for the investigation of the fundamental properties of molecular crystals. Doping, for example, allowed to significantly improve the performance of organic devices \cite{Walzer2007,Pfeiffer1998}, and moreover, lead to fascinating and sometimes unexpected observations. One of the most remarkable examples are the fullerides, doped fullerene compounds. Their properties range from band or Mott insulators to metallic and superconducting behaviors as a function of doping level and crystal structure \cite{Gunnarsson1997,Forro2001}. Other examples comprise a transition from a Luttinger liquid to a Fermi liquid in potassium doped carbon nanotubes \cite{Rauf2004}, metallic phases in metal doped 3,4:9,10-perylenetetracarboxylic dianhydride (PTCDA) \cite{Shklover2001,Craciun2006}, or predictions of quasi-one-dimensional K-O chains in potassium doped PTCDA thin films \cite{Zazza2007}. More recently, interesting phenomena were observed in metal doped, hydrocarbon based molecular materials. It has been observed that some molecular crystals consisting of polycyclic aromatic hydrocarbons demonstrate superconductivity upon alkali metal addition. The associated transition temperatures (T$_c$) into the superconducting state have been reported to be rather high (for instance T$_c$ = 18\,K for K$_3$picene) \cite{Mitsuhashi2010}.

\par

In this contribution we review recent investigations of the electronic excitations of two representative phenantrenes, chrysene and picene. We have applied electron energy-loss spectroscopy (EELS) in order to determine the electronic excitation spectrum in the undoped as well as the electron doped case. Doping with electrons was achieved via the addition of potassium and calcium, respectively.

\section{Experimental}

Our EELS investigations require thin samples with a thickness of about 100\,nm only. In this contribution we show results of both single crystalline and polycrystalline samples, respectively. On the one hand, thin films of single crystals were cut from a macroscopic single crystal platelet with the help of an ultramicrotome using a diamond knife. Previous to the thin film preparation, single crystals of very good quality were prepared via physical vapor growth in a vertical geometry. Chrysene as well as picene were sublimed from a glass surface and the crystal growth occurred on a Al foil on top. The growth lasted 12 hours and resulted in thin, singly-crystalline platelets with typical dimensions of about 0.5\,mm x 0.5\,mm.

\par On the other hand, also large thin films of the organic compounds have been produced by thermal evaporation under high vacuum onto a single crystalline substrates (KBr) kept at room temperature in a separate vacuum chamber. During the vacuum deposition the film thickness was monitored \emph{in-situ} via a quartz crystal microbalance. Subsequent to the evaporation, the films are floated off in destilled water, mounted onto standard electron microscopy grids \cite{Fink1994}, incorporated into an EELS sample holder, and transferred into the EELS spectrometer \cite{Fink1989,Roth2014}.

\par

Prior to the spectroscopic investigations all samples were charaterized \emph{in-situ} using electron diffraction. Furthermore, in case of the single crystalline samples the diffraction experiments were used to orient the momentum transfer parallel to selected reciprocal lattice directions.

\par

In addition, we intercalated the samples with potassium and calcium in a preparation chamber (base pressure lower than 10$^{-10}$\,mbar) of the EELS spectrometer. In detail, the samples have been exposed to a metal vapor in a distance of 30\,mm directly over an alkali metal getter sources (SAES GETTER S.p.A Viale Italia 77, 20020 Laina, Italien) or over a heated molybdenum crucible containing calcium. This doping procedure has been carried out in several steps until saturation, i.\,e., until no further stoichiometric changes could be observed in the spectra. To be able to further analyze the doping induced changes, we additionally measured the C\,1$s$ and K\,2$p$ or Ca\,2$p$ core excitation edges in order to determine the amount of potassium or calcium in our doped samples. We compared the associated core-level excitation intensities with those of other doped molecular films with well known stoichiometry, i.\,e., K doped C$_{60}$ \cite{Knupfer2001}, or analyzed the intensities in a quantitative manner. A detailed description of these procedures can be found, e.\,g., in \cite{Roth_K3picene2011,Roth_Ca2013}.

\par

All measurements were carried out using a 172\,keV spectrometer thoroughly discussed in detail in previous publications \cite{Fink1989,Roth2014}. Note that at this high primary beam energy only singlet excitations are possible The energy and momentum resolution is 85\,meV and
0.03\,\AA$^{-1}$ for all measurements, respectively. The EELS signal, i.\,e., the loss function Im[-1/$\epsilon(\textbf{q},\omega)$], which is proportional to the dynamic structure factor S($\textbf{q},\omega$), was determined for various momentum transfers, $q$, parallel to the film surface [$\epsilon(\textbf{q},\omega)$ is the dielectric function]. The ability of EELS to measure as a function of momentum transfer can give valuable insight into the nature and dynamics of different excitations such as, e.\,g., plasmons or excitons \cite{Roth2014,Schuster2007,Knupfer1999,Kramberger2008}.

\section{Results and Discussion}
\subsection{EELS On Pristine Aromatic Hydrocarbon Systems}

\begin{figure*}[t]
\includegraphics[width=0.48\linewidth]{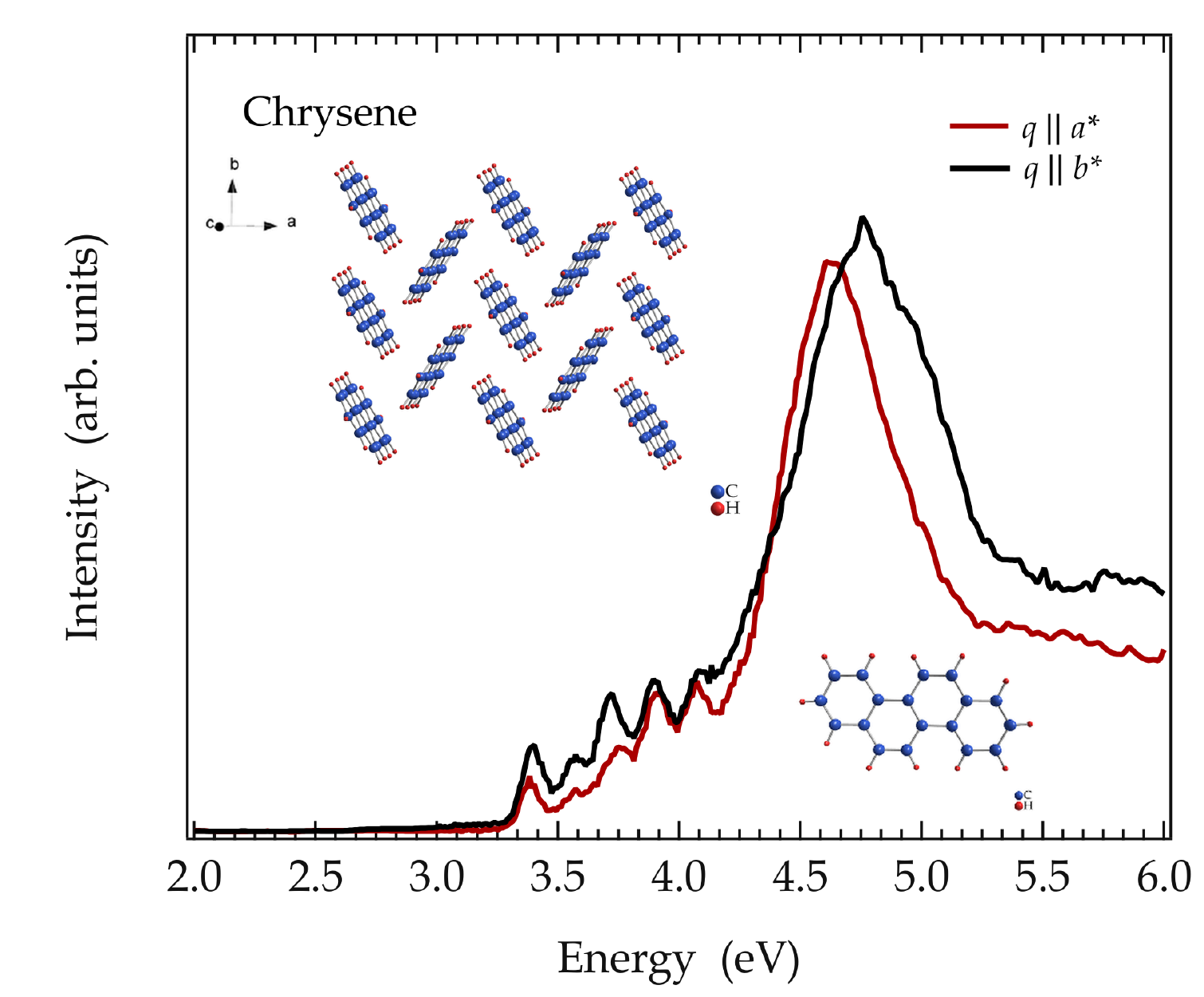}
\includegraphics[width=0.48\linewidth]{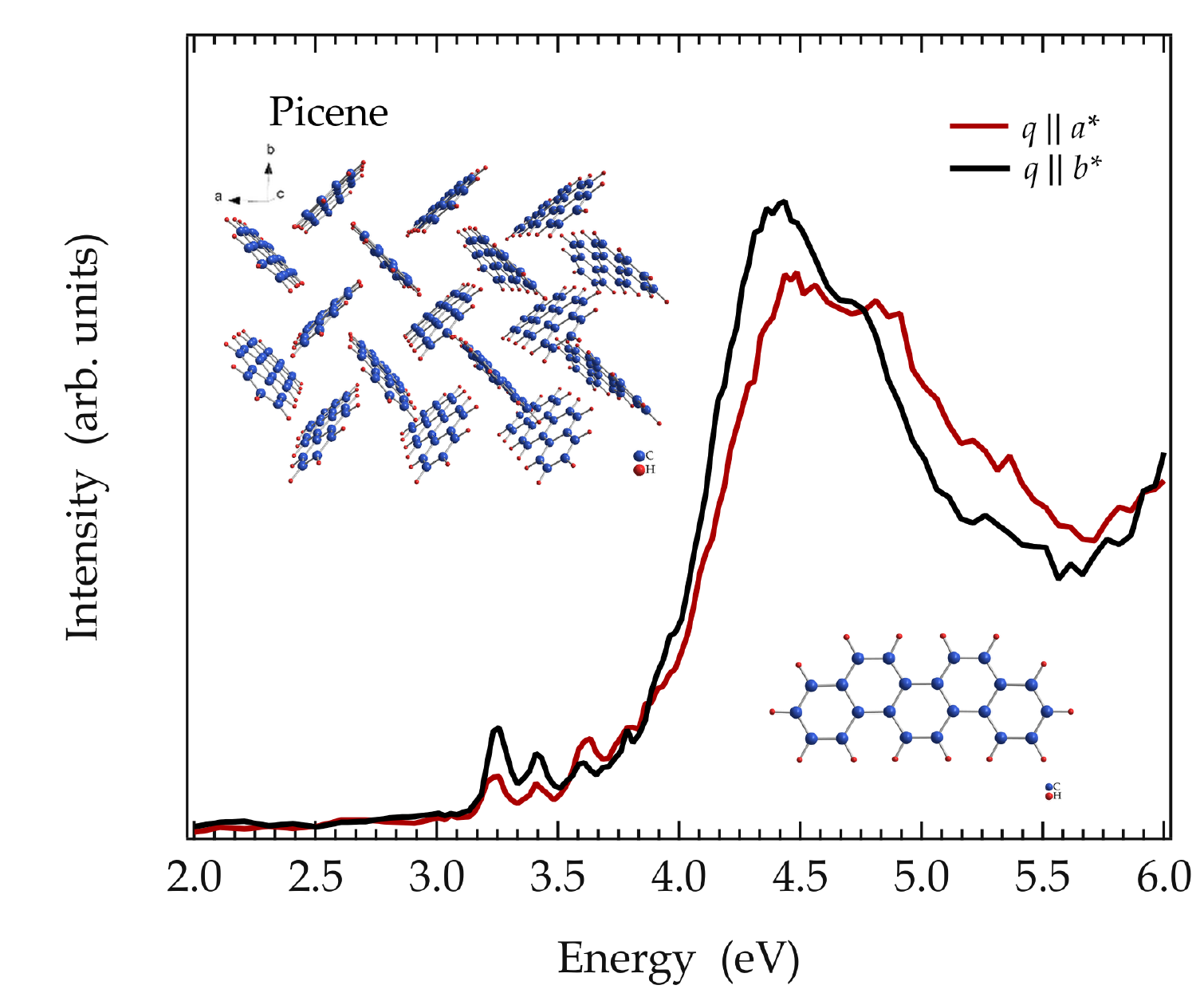}
\caption{Loss function of chrysene (left panel) and picene (right panel) single crystals for momentum transfers parallel to the two main directions in reciprocal space, $a^*$ and $b^*$. The absolute value of the momentum transfer is 0.1\,\AA$^{-1}$ for all measurements. Insets show the schematic representation of the molecular structure of the investigated compounds as well as the crystal structure in real space---the so-called herringbone structure, which represents the typical arrangement of such organic molecules.}
\label{fig1}
\end{figure*}

We start the discussion of the electronic excitation spectra with those of the pristine, single-crystalline hydrocarbons chrysene and picene. Fig.\,\ref{fig1} presents the loss function of these two materials for momentum transfer directions parallel to the $a^*$ and $b^*$ reciprocal lattice directions. The experimental data presented in Fig.\,\ref{fig1} are taken with a small value of the momentum transfer $q$ of 0.1\,\AA$^{-1}$, which represents the so-called optical limit. Taking into account the anisotropic molecular and crystal structure of chrysene and picene, it is reasonable to expect an anisotropic electronic response for these materials as well. Fig.\,\ref{fig1} depicts that the electronic excitation spectrum within the reciprocal $a^*$,$b^*$-plane indeed is anisotropic, in particular in terms of the different excitation intensities of the spectral features. In general, the well-structured spectra with several clear maxima  are a signature of the energetically sharp and well-defined molecular electronic levels of chrysene and picene, which are due to excitations between occupied and unoccupied $\pi$-derived electronic levels and which remain relatively unchanged going to the solid state. The data as shown in Fig.\,\ref{fig1} are in very good agreement to those from polycrystalline thin films of the two hydrocarbons \cite{Roth_picene2011,Roth_chrysene2012}. Moreover, for picene they also have been calculated using state-of-the-art many-body calculations based on the Bethe–Salpeter equation \cite{Roth_hydrocarbons2013}. The main features of the spectra are well reproduced by the calculations, which can therefore safely be used to interpret the experiments.

\par

The excitation onset in the spectra as shown in Fig.\,\ref{fig1} represents the so-called optical gap of the two materials, which is at 3.3\,eV for chrysene and at 3.15\,eV for picene, respectively. In both cases, the excitation onset is followed by additional well-separated features at 3.41\,eV, 3.57\,eV, 3.72\,eV, 3.90\,eV, and 4.08\,eV for chrysene and at 3.25, 3.41, 3.61, 3.77, and 3.93\,eV for picene. The main features in our excitation spectrum are in very good agreement with optical absorption data \cite{Gallivan1969,Tanaka1965,Fanetti2012}. In general, the energetically lowest lying electronic excitations in organic molecular solids often are excitons, i.\,e., bound electron-hole pairs \cite{Pope1999,Knupfer2003,Hill2000}. The criterion that has to be considered in order to analyze the excitonic character and binding energy of an excitation is the energy of the excitation with respect to the so-called transport energy gap, which represents the energy needed to create an unbound, independent electron-hole pair. This transport energy gap has been estimated previously to be about 4.2\,eV for chrysene \cite{Sato1987} and 4.05\,eV for picene \cite{Sato1987,Roth2010}. Consequently, the excitation features in pristine chrysene and picene which are observed below this value for the transport energy gap can be attributed to singlet excitons and the exciton binding energy of the lowest lying exciton is as large as about 0.9\,eV.

\par

\begin{figure*}[t]
\includegraphics[width=0.48\linewidth]{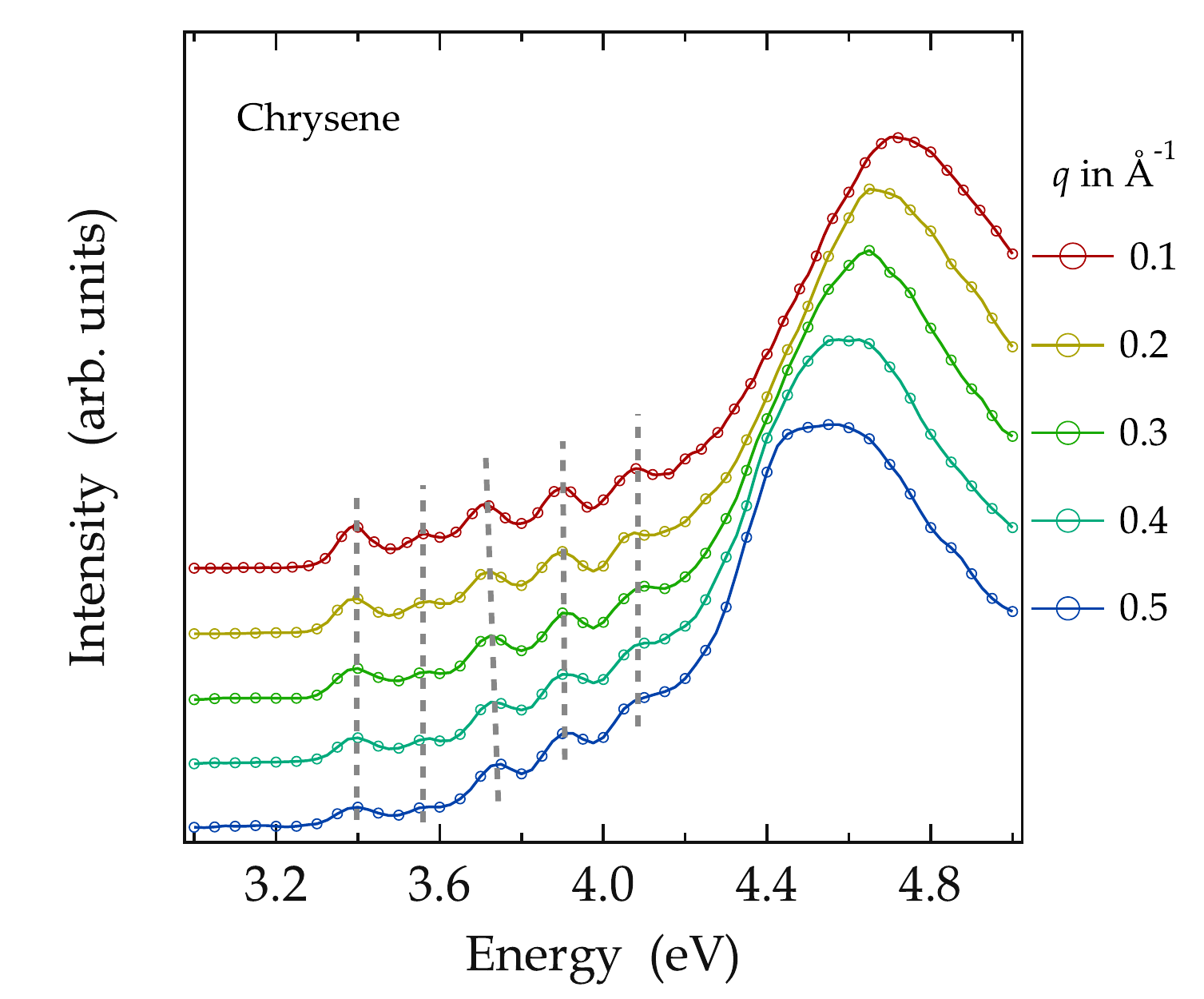}
\includegraphics[width=0.48\linewidth]{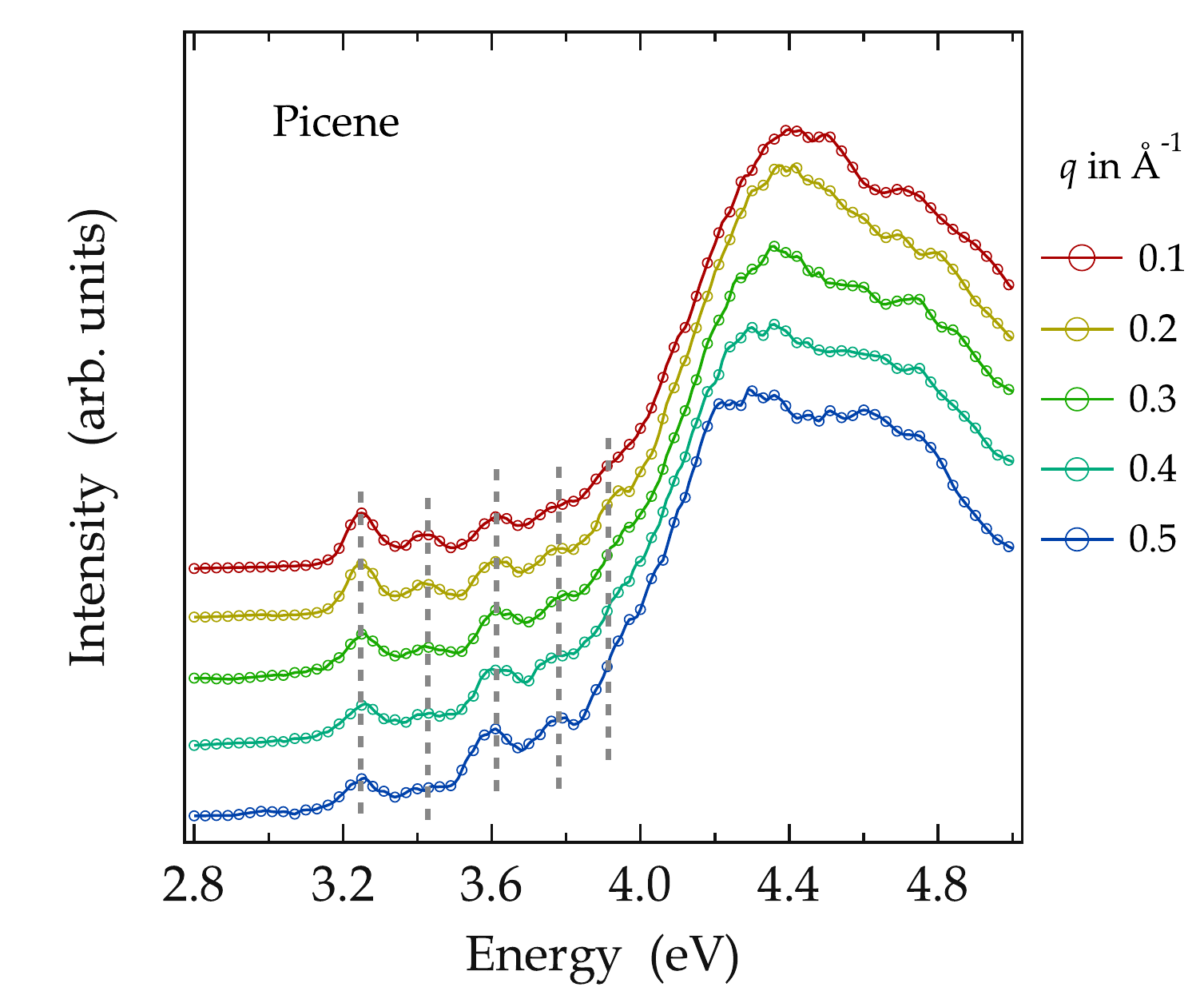}
\caption{Momentum dependence of the EELS spectra of solid (polycrystalline) chrysene (left panel) and picene (right panel). The measurements were carried out at T\,=\,20\,K. The grey lines are intended as a guide to the eye.}
\label{fig2}
\end{figure*}

We have also determined the electronic excitation spectra with increasing momentum transfer $q$ as shown in Fig.\,\ref{fig2}. In chrysene, all identified excitons do not change in energy within a momentum range up to 0.5\,\AA$^{-1}$. Consequently, in the framework of an exciton band-structure description, this is equivalent to a vanishing group velocity, in other words the excitons are rather localized. This is very similar to the observations for picene (see right panel of Fig.\,\ref{fig2}), however, there are also differences in the momentum dependence of the excitation spectra between chrysene and picene. For picene, we observe that one of the low lying excitations (at 3.61\,eV) increases in spectral weight with increasing momentum transfer, which can be interpreted as evidence for the additional contribution of a charge transfer exciton to the spectra \cite{Roth_picene2011}.

We note that the observation of non-dispersing, localized excitons in chrysene and picene is in significant contrast to what was seen for tetracene and pentacene (formed by four and five benzene rings fused in a linear manner, respectively), where the electronic excitation across the energy gap has substantially more relative spectral weight and shows a complex and anisotropic dispersion \cite{Schuster2007,Roth_pentacene2012,Roth_hydrocarbons2013}. This difference has been attributed to the different nature of the lowest excitations in phenantrenes (chrysene and picene) and acenes (tetracene and pentacene), respectively. While, e.\,g., for pentacene the excitation onset is predominantly due to a charge-transfer exciton, in picene it is related to a strongly localized Frenkel exciton \cite{Cudazzo2012,Cudazzo2013}. It has been argued, that the subtle interplay of direct and exchange-type electron-hole interactions, as well as of the transfer integrals (hopping) in the molecular solid are responsible for such a significant variation in regard of the exciton ground state  \cite{Cudazzo2012,Cudazzo2013}, which is a further nice example of the diversity of the physical properties of molecular solids.

\subsection{EELS On Doped Aromatic Hydrocarbon Systems}

\subsubsection{Potassium Intercalation}

The addition of alkali metals, here potassium, to organic thin films results in a significant change of the electronic properties of these films. This is caused by the relatively easy diffusion of the potassium atoms into the organic material and the concurrent charge transfer of the outer 4\,$s$ electron to the organic molecules. In a first step, we have always characterized the structure of the films using in-situ electron diffraction. This demonstrated a polycrystalline structure of the films \cite{Roth_K3picene2011,Roth_PhD}, which in the case of K doped picene agrees well with data from x-ray diffraction \cite{Mitsuhashi2010}. In other words, the films consist of crystalline grains that are larger than the wave-length of the studied plasmons.

\begin{figure}[h]
\centering
\includegraphics[width=0.7\linewidth]{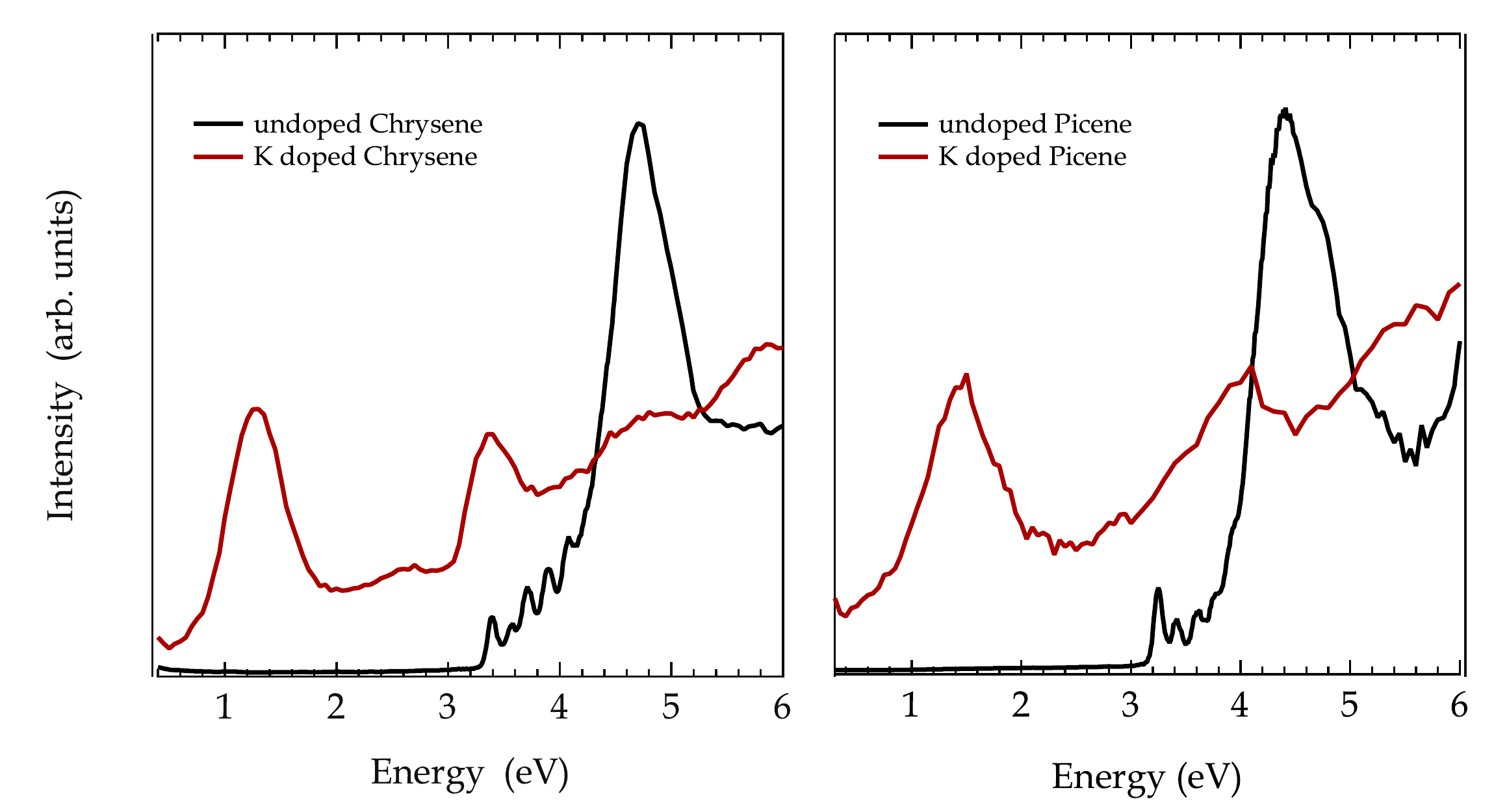}
\caption{Comparison between the loss function for undoped and K intercalated chrysene (left panel) and picene (right panel).}
\label{fig3}
\end{figure}

Fig.\,\ref{fig3} illustrates the consequences of the doping induced charge transfer for the electronic excitations in chrysene and picene. For both materials, we have determined a composition of about three potassium ions per hydrocarbon molecule \cite{Roth_chrysene2012,Roth_K3picene2011}, and the addition of potassium results in the appearance of a prominent excitation feature, which is observed at about 1.3\,eV for K$_3$chrysene and at about 1.5\,eV for K$_3$picene.  Further, the doping process leads to a broadening of the excitation features at higher energies and to a downshift of some of these excitations. The latter can be rationalized by a relaxation of the molecular structure and electronic levels upon addition of electrons to anti-bonding $\pi$* orbitals. The observed broadening of the excitation spectra might be due to a symmetry lowering of the molecules upon charging as well as to a reduced life-time of the excitations in the doped systems.

\par

\begin{figure}[h]
\centering
\includegraphics[width=\linewidth]{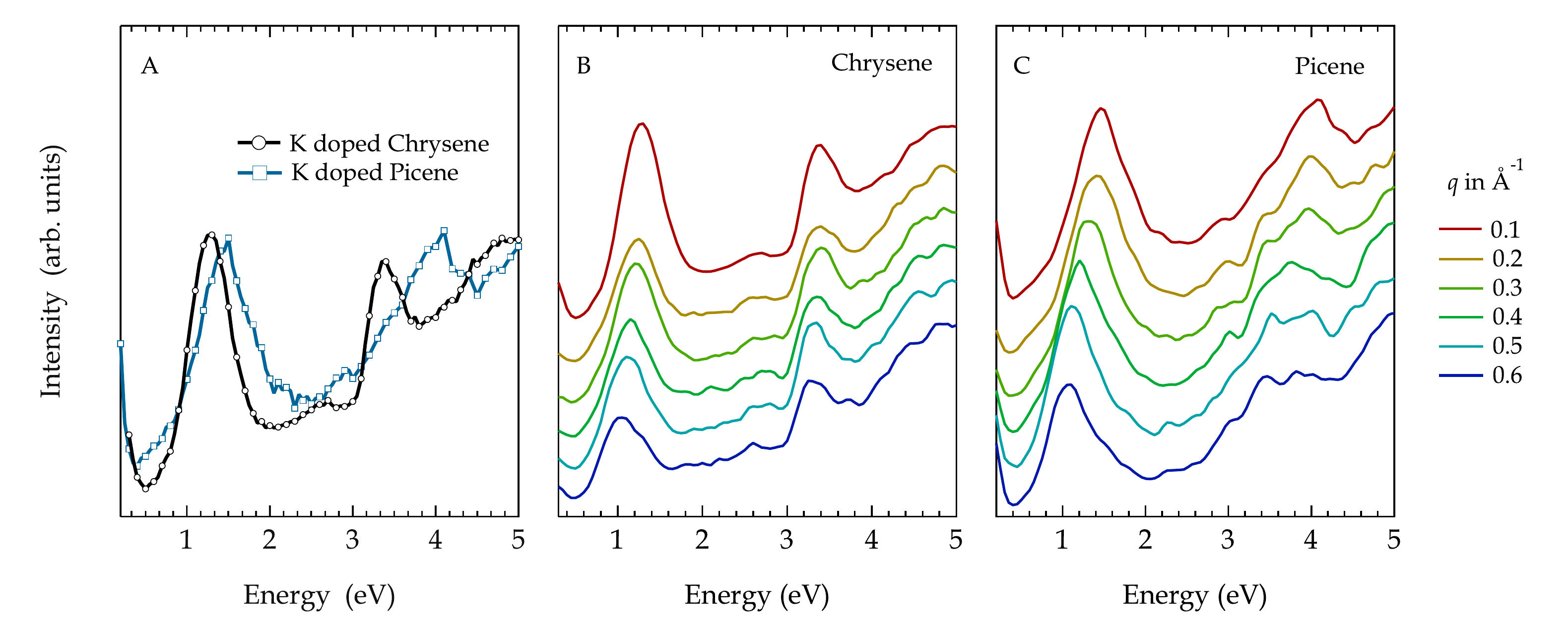}
\caption{A: Comparison between the loss function in the range of 0 - 5\,eV for potassium intercalated chrysene (black circles) and picene (blue squares). Momentum dependence of the EELS spectra of K intercalated chrysene (B) and picene (C). $q$ is increasing from top to bottom spectra.}
\label{fig4}
\end{figure}

The close similarity of the electronic excitations in chrysene and picene is true for the undoped as well as for the doped systems. This is further illustrated in Fig.\,\ref{fig4} where we show a direct comparison of the excitation spectra for potassium doped chrysene and picene. From panel A it becomes clear that the low energy region of the corresponding loss spectra is characterized by three spectral features, which for doped chrysene appear at somewhat lower energies. This close similarity has also been underlined by a Kramers-Kronig analysis of these data. Interestingly, it was consistently possible to carry out this analysis under the assumption of a metallic ground state for potassium doped chrysene and picene \cite{Roth_chrysene2012,Roth_K3picene2011}. Also, the measured excitation spectra agree well with those from state-of-the-art calculations for a metallic K$_3$picene phase \cite{Cudazzo2011}. This would be in agreement with the reported superconductivity for potassium doped picene \cite{Mitsuhashi2010} but is in contrast to recent photoemission studies of doped picene systems, which have revealed an insulating ground state that was ascribed to strong electronic correlations and a Mott-insulating ground state \cite{Mahns2012,Caputo2012,Ruff2013}. Moreover, there is also evidence from theoretical considerations that doped hydrocarbon crystals, when doped with three potassium atoms per molecule, are close to a metal-insulator transition into a Mott insulating phase \cite{Kim2012,Giovannetti2011,Ruff2013}. These contradictory experimental and theoretical findings have not been resolved yet. Differences in various experiments might arise from the unknown phase behavior and phase diagram of potassium doped picene. A detailed knowledge of stable doped phases and their characteristics would lay the ground for further investigations of the electronic properties \cite{Naghavi2014,Heguri2014}. This is particularly emphasized by the exploration and rationalization of the electronic properties of the fullerides in the past. There, the determination of the respective phase diagrams have been an indispensable ingredient \cite{Gunnarsson1997,Forro2001,Poirier1995,Benning1993,Knupfer1994_K_C70}.

\par

\begin{figure}[h]
\centering
\includegraphics[width=0.48\linewidth]{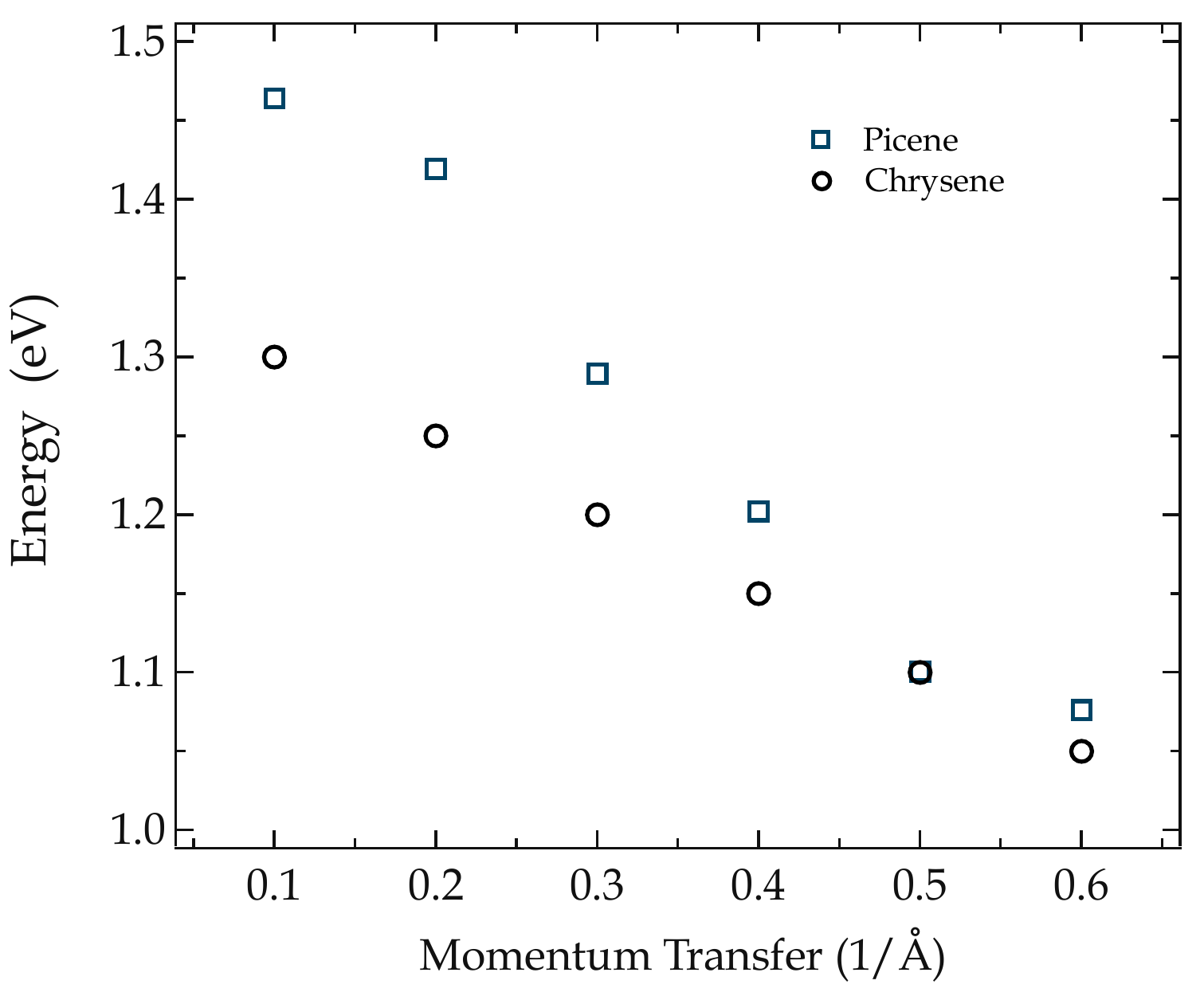}
\caption{Plasmon dispersion for K doped chrysene in comparison to K doped picene up to a momentum transfer $q$ of 0.6\,\AA$^{-1}$.}
\label{fig5}
\end{figure}

Taking into account the above mentioned Kramers-Kronig analysis, which yields results consistent with a metallic state of K$_3$chrysene and K$_3$picene, the prominent excitation feature at around 1.5\,eV for both compounds represents the charge carrier plasmon. The dispersion of this charge carrier plasmon in potassium doped chrysene and picene is summarized in Fig.\,\ref{fig4}, panels B and C. In both cases, the plasmon dispersion is negative (see also Fig.\,\ref{fig5}), which is in contrast to the traditional picture of metals with a homogeneous electron gas. In these models a quadratic and positive plasmon dispersion is expected, whereas the slope of this dispersion is proportional to the mean squared Fermi velocity.

\par

On the other hand, molecular solids, even if they become metallic, differ substantially from conventional metals. The electronic wave function which form the conduction bands are strongly localized on the molecules. This causes also a momentum dependence of the loss function (and the dielectric function) which is quite different from that of a simple metal. This competition between charge localization and metallicity can result in a negative plasmon dispersion in the (metallic) molecular solid \cite{Cudazzo2011}. It is interesting to note that investigations of the plasmon dispersion in K$_3$C$_{60}$ molecular solids have also revealed an unusual behavior, the plasmon dispersion is characterized by a vanishing momentum dependence \cite{Gunnarsson1996}.

\subsubsection{Calcium Intercalation}

\begin{figure*}[t]
\includegraphics[width=0.48\linewidth]{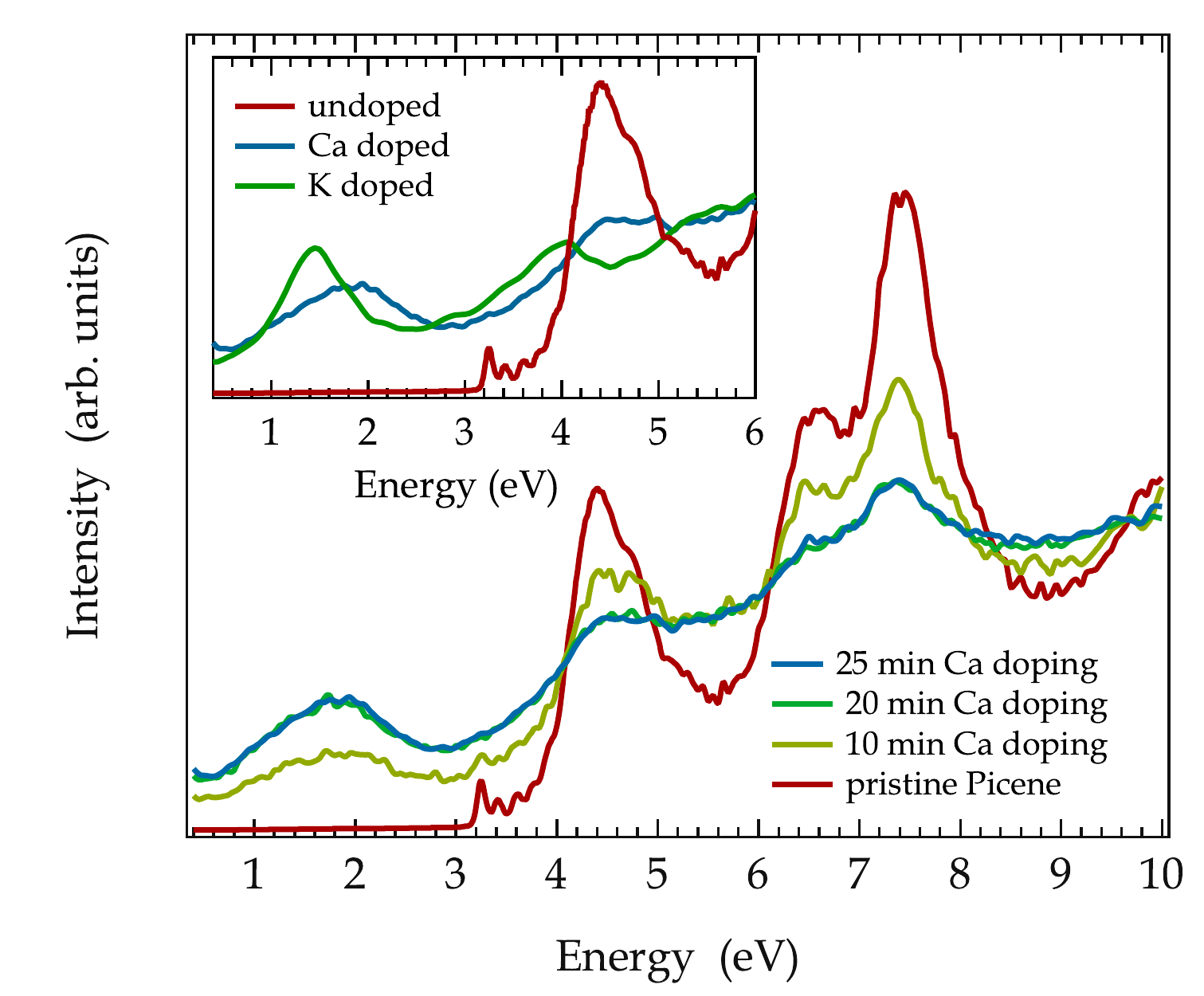}
\includegraphics[width=0.48\linewidth]{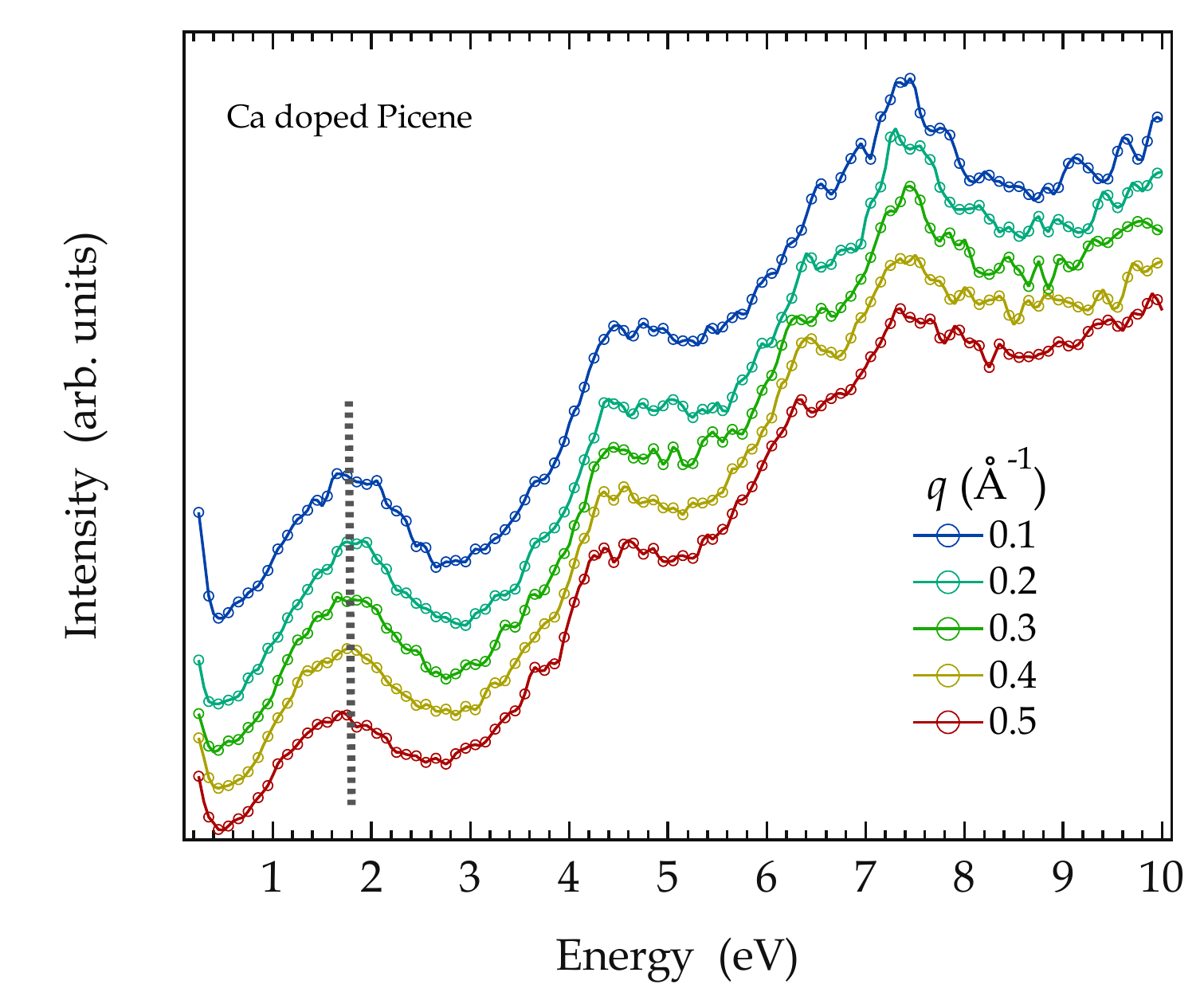}
\caption{Left panel: Loss function of picene as a function of evaporation time of Ca. Inset: Comparison between pristine, K doped and Ca doped picene in an energy range up to 6\,eV. Right panel: Momentum dependence of the EELS spectra of Ca doped picene in the range between 0.1 - 0.5\,\AA$^{-1}$. The grey line is intended as a guide to the eye.}
\label{fig6}
\end{figure*}

In contrast to alkali metals, alkaline-earth elements are divalent, and thus in principle offer the opportunity to achieve higher doping levels when added to organic materials. It is therefore interesting to investigate the influence of the addition of alkaline-earth elements, such as calcium, to the hydrocarbons discussed in this article. In the following we present our results for Ca-doped picene films.

\par

In Fig.\,\ref{fig6} (left panel) we show the evolution of the excitation spectra upon Ca addition. It becomes clear that the addition of the metal again results in a broadening of all spectral features which are representative of picene. Also, we observe an extra excitation at lower energies, i.e. below the optical gap of picene. For Ca addition the feature is found at about 1.9\,eV. Fig.\,\ref{fig6} also illustrates how different amounts of Ca influence the spectra. They change in a simple manner that suggests the formation of a single Ca-intercalated picene phase up to the saturation level that can be achieved under the experimental condition applied here. This saturation level corresponds to a composition of Ca$_3$picene, i.\,e., again 3 metal atoms per molecule are incorporated in this molecular material, which had been analyzed by core level spectroscopy \cite{Roth_Ca2013}. Importantly, these core level data also demonstrated that there is the formation of hybrid orbitals between picene and Ca-derived states and that the effective charge transfer per Ca is one electron only, in contrast to the expectation of a higher doping level per added Ca.

\par

The formation of hybrid states most likely is also responsible for the relatively large width of the doping induced spectral feature at 1.9\,eV. It is much broader than that observed for potassium doping, which can be seen from the inset of Fig.\,\ref{fig6} where we show a direct comparison of the respective data. Also, the Kramers-Kronig analysis of the measured loss function indicated that the ground state of Ca-doped picene is non-metallic with a rather small gap in the electronic spectrum \cite{Roth_Ca2013}.

\par

The fact that Ca doped picene is not a purely ionic compound but characterized by hybrid states between Ca and the $\pi$ electronic system of picene parallels the situation that has been reported for other Ca doped $\pi$ conjugated materials, Ca-doped C$_{60}$ and Ca-intercalated graphite CaC$_6$. In these cases, the hybridization is responsible for a reduction of the charge that is transferred to the molecules or graphite planes in
comparison to the value of two transferred electrons in a purely ionic picture \cite{Saito1992,Chen1992,Romberg1994}. In addition, such hybridization effects have also been reported for $\pi$ conjugated systems doped with other alkaline-earths, such as Ba-doped C$_{60}$ or Ba-doped single wall carbon nanotube (SWCNTs) \cite{Saito1993,Erwin1993,Knupfer1994,Liu2004}.

\par

Finally, we present the momentum dependence of the lowest electronic excitation of Ca-doped picene in Fig.\,\ref{fig6} (right panel). The evolution of these data upon increasing momentum transfer shows a rather momentum independent behavior. Thus, the underlying electronic excitations can be regarded as being localized, quite in contrast to the case of K$_3$picene discussed above. These observations support  the discussed differences in the electronic structure between Ca$_3$picene and K$_3$picene.

\section{Summary}

To summarize, we investigated the electronic excitations of chrysene and picene using electron energy-loss spectroscopy. We have identified the excitonic character and the rather localized wave function of the excitations across the band gap. The addition of potassium results in a substantial change in the excitation spectra for both chrysene and picene. A new low energy feature is induced which can be interpreted as a charge carrier plasmon with a surprisingly negative dispersion. This interpretation, however is in contrast to photoemission results which do not find a metallic, potassium doped picene phase. Further studies are required to solve this discrepancy. Ca addition to picene also results in a new low energy excitation. In this case, however, this excitation does not show a dispersion and is much broader as compared to the potassium intercalation. Taking into account previous core level data, these findings can be rationalized in terms of the formation of hybrid states between calcium and picene, a scenario which, e.\,g., resembles that in Ca intercalated graphite.

\begin{acknowledgments}
We thank M. Naumann, R. Sch\"onfelder, R. H\"ubel and S. Leger for technical assistance. Part of this work has been supported by the Deutsche Forschungsgemeinschaft under KN393/13, KN393/14.
\end{acknowledgments}

\end{document}